\pdfoutput=1
\documentclass[pre,onecolumn,12pt,aps,superscriptaddress,citeautoscript,tightenlines,nofootinbib]{revtex4-2}

\usepackage[dvipsnames]{xcolor}
\usepackage{amsmath,amssymb}
\usepackage{graphicx}
\usepackage{footmisc}
\usepackage{bbold,soul}
\usepackage{lipsum}
\usepackage{enumitem}
\usepackage{hyperref}
\usepackage{physics}
\usepackage{gensymb}
\usepackage{watermark}
\usepackage{mathtools}

\usepackage{mc}
\let\vec\mathbf

\begin{document}

\title{Emulating Expert Insight: \\ A Robust Strategy for Optimal Experimental Design}

\author{Matthew R.~Carbone}
\email{mcarbone@bnl.gov}
\affiliation{Computational Science Initiative, Brookhaven National Laboratory, Upton, NY 11973, USA}

\author{Hyeong Jin Kim}
\affiliation{Center for Functional Nanomaterials, Brookhaven National Laboratory, Upton, NY 11973, USA}

\author{Chandima Fernando}
\affiliation{National Synchroton Light Source II, Brookhaven National Laboratory, Upton, NY 11973, USA}

\author{Shinjae Yoo}
\affiliation{Computational Science Initiative, Brookhaven National Laboratory, Upton, NY 11973, USA}

\author{Daniel Olds}
\affiliation{National Synchroton Light Source II, Brookhaven National Laboratory, Upton, NY 11973, USA}

\author{Howie Joress}
\thanks{orcid.org/0000-0002-6552-2972}
\affiliation{Materials Measurement Science Division, National Institute of Standards and Technology, Gaithersburg, MD 20899, USA}

\author{Brian DeCost}
\thanks{orcid.org/0000-0002-3459-5888}
\affiliation{Materials Measurement Science Division, National Institute of Standards and Technology, Gaithersburg, MD 20899, USA}

\author{Bruce Ravel}
\thanks{orcid.org/0000-0002-4126-872X}
\affiliation{Materials Measurement Science Division, National Institute of Standards and Technology, Gaithersburg, MD 20899, USA}

\author{Yugang Zhang}
\email{yuzhang@bnl.gov}
\affiliation{Center for Functional Nanomaterials, Brookhaven National Laboratory, Upton, NY 11973, USA}

\author{Phillip M.~Maffettone}
\email{pmaffetto@bnl.gov}
\affiliation{National Synchroton Light Source II, Brookhaven National Laboratory, Upton, NY 11973, USA}

\date{\today}

\begin{abstract}
The challenge of optimal design of experiments (DOE) pervades materials science, physics, chemistry, and biology. 
Bayesian optimization has been used to address this challenge in vast sample spaces, although it requires framing experimental campaigns through the lens of maximizing some observable.
This framing is insufficient for epistemic research goals that seek to comprehensively analyze a sample space, without an explicit scalar objective (e.g., the characterization of a wafer or sample library). 
In this work, we propose a flexible formulation of scientific value that recasts a dataset of input conditions and higher-dimensional observable data into a continuous, scalar  metric.
Intuitively, the scientific value function  measures where observables change significantly, emulating the perspective of experts driving an experiment, and can be used in collaborative analysis tools or as an objective for optimization techniques.
We demonstrate this technique by exploring simulated phase boundaries from different observables, autonomously driving a variable temperature measurement of a ferroelectric material, and providing feedback from a nanoparticle synthesis campaign. 
The method is seamlessly compatible with existing optimization tools, can be extended to multi-modal and multi-fidelity experiments, and can integrate existing models of an experimental system. 
Because of its flexibility, it can be deployed in a range of experimental settings for autonomous or accelerated experiments. 
\end{abstract}

\maketitle

\section{Introduction}
The combination of  automation and artificial intelligence (AI) to create closed-loop self driving, autonomous laboratories---or human-interfaced acceleration platforms---has begun revolutionizing scientific research across chemistry~\cite{hase2019next, escalate}, biology~\cite{senior2020improved, Narayanan_2021}, and materials science~\cite{Stach_2021, campbell2021outlook,barbour2022advancing,konstantinova2022machine,seifrid2022reaching}.
These contemporary platforms mostly use single AI agents, but can also leverage the added value of multiple agents working in tandem~\cite{maffettone2023self}.
To date, most efforts in agent development have focused on designing algorithms that optimize a target value~\cite{Noack_2021} or resource allocation~\cite{gamification,mcdannald2022fly}.
Unfortunately, these approaches to agent design do not encompass the research settings where the objective is more epistemic; that is, the research pertains to comprehensive understanding or interpretation of an experimental space~\cite{mcdannald2022reproducible}, and not the optimization of a target. 
Examples of epistemic objectives are ubiquitous in characterization~\cite{MaffettoneMASS}, user facilities, and ``science as a service'' platforms~\cite{li2020autonomous}.
These bring forth a new challenge in how to leverage AI advancements for  optimal experiment design.
\par
Research motivated by comprehensive understanding of a system is common across disciplines. 
It appears  in problems involving spatial characterization or fixed sample spaces, such as device mapping~\cite{Hua_2021, Olds_2020}, tomography~\cite{barbour2022advancing}, or phase mapping~\cite{Kusne_2020,joress2020high}. 
It also is recurrent when searching large plateaus of space for sharp changes, such as in searching for reactivity~\cite{Caramelli_2021} or phase changes
~\cite{Maffettone_XCA, joress2020high}.
Nonetheless, when research questions are directed more by understanding a system than by optimizing the system for a single property, certain measurements will still prove more valuable than others. 
Traditional DOE approaches the exploration of a known space to explain the variation of a response function in that space, albeit it is not adaptive and expects each input parameter to impact the response function~\cite{StatisticsForExperimenters}.
In the case of allocating limited resources over independent samples, reinforcement learning (RL) has been used to extract maximal value~\cite{gamification}.
When a model for the system is available Bayesian inference can be used to query data that will best mitigate the uncertainty of that model \cite{konstantinova2022_PRR,koplik2022topological,Kusne_2023_Matter}.
Bayesian optimization over expected information gain~\cite{balandat2020botorch} can be used to explore the experimental space; however, without a conversion between an observable and a finite objective, Bayesian optimization cannot be effectively leveraged. 
\par
With an epistemic goal, an optimal agent will therefore yield an experimental design that produces the best dataset for understanding the experimental space.
This understanding would be derived from expert interpretation, modeling, AI, or some combination of techniques. 
Furthermore, the agent should be able to operate with or without a model of space or the observable. 
It must also be robust to the ``cold start" problem~\cite{swersky2013multi}, operating efficiently under initially extremely data- and information-limited conditions.
Lastly, it would be beneficial for any agent to make use of contemporary advancements in optimization methods. 
\par
Herein, we propose a generic scientific value function (SVF) that recasts a dataset of observables into scalar measures of ``value" by mirroring the perspective and actions of human experts.
While natively model-free, the SVF can incorporate models of the experimental or observable space.
Crucially, it can be used as an optimization target in other procedures, such as Bayesian or Monte Carlo optimization~\cite{balandat2020botorch}.
We demonstrate the application of the SVF through: i) a simulated X-ray diffraction (XRD) phase mapping of first- and  second-order transitions; ii) a simulated absorption spectroscopy study of a periodic phase boundary; iii) a variable temperature X-ray total scattering study of BaTiO$_3$; and iv) an ultraviolet-visible (UV-vis) absorption spectroscopic analysis of nanoparticle synthesis conditions. 
The adaptive applications of the SVF are accomplished by optimizing the next measurement value over experimental space using a Gaussian process~\cite{Rasmussen2006GP} surrogate model and Bayesian optimization~\cite{shahriari2015taking,frazier2018tutorial,balandat2020botorch}. 
This work creates opportunities for optimal dataset creation and research acceleration without a pre-existing optimization target, and will find broad applicability across scientific disciplines. 

\section{Results \& Discussion}

\subsection{A surrogate function for scientific value} \label{subsec:a surrogate function for scientific value}
We set out to construct a surrogate function for scientific value that would emulate the judgement of an expert scientist without a model for their experimental system. 
Consider a prototypical example of mapping the phase diagram of a material over multiple dimensions.
A rational scientific goal would be to measure every unique phase at least once, and measure with greater resolution across phase boundaries. 
At the start of a campaign, the measurement of every location has the same potential value. 
As the campaign progresses, measuring the regions where the observable is not changing with the ordinate becomes less valuable than measuring regions of rapid change. We defined the Scientific Value Function (SVF), $U,$ to capture this intuition.

\par
First, we consider an input space $\mathcal{X},$ where queries of $\vec x_i \in \mathcal{X}$ comprise a dataset $\mathcal{D}_N \coloneqq \{ (\vec x_1, \vec y_1), ... (\vec x_N, \vec y_N)\}$,
where $\vec y_i$ are noisy, multi-dimensional observations of some function, $\vecf(\vec x_i$), such as a diffraction image.
We further define two correlation functions for both the input space and the observation space, $h(\vec x_i, \vec x_j)$ and $g(\vec y_i, \vec y_j)$, respectively.
The default correlation function used in this work for both $h$ and $g$ is the Euclidean distance, or $L_2$ norm, $\norm{\cdot}_{L_2}$.
Thus, we define the dataset-dependent Scientific Value Function (SVF) as
\begin{equation}\label{eq:U(x,D)}
  U(\vecx_i, \vecy_i | \mathcal{D}_N) = \sum_{(\vec x_j, \vec y_j) \in \mathcal{D}_N}
  g(\vec y_i, \vec y_j)
  \exp{ - \frac{1}{2}\frac{h(\vec x_i, \vec x_j)^2}{
      h_{\mathrm{min}}(\vec x_i | \mathcal{D}_N)^2}
  },
\end{equation}
where $h_\mathrm{min}(\vecx_i | \mathcal{D}_N)$ is the distance between $\vecx_i$ and its nearest neighbor in $\mathcal{D}_N.$ Using Eq.~\eqref{eq:U(x,D)}, the scientific value can be computed for all inputs in $\mathcal{D}_N.$ 
\par

The SVF considers the individual value of a new datum with respect to each existing datum and sums over all members of the dataset for a net value. 
The first term considers where the observable is distinct from those contained in the dataset, and thus valuable. 
The second term decays that value with respect to how far the data are in the input space.
In order to avoid overestimating the value of local regions, the second term is regularized by the nearest neighbors of points in input space.
We considered other forms of the SVF that would use these correlation functions (\emph{e.g.,} proportionate or derivative-like functions); however, chose the form of the second term such that it would be limited between 1 and 0, and had options for regularization. 
\par

This formalism offers a few key features and advantages.
Firstly, it adequately reflects the intuition of researchers in practice.
It also reduces the dimensionality of the observable space to a scalar objective function that can be readily optimized.
While the approach is natively model free, the correlation functions $g$ and $h$ are flexible and can readily incorporate models of the system.
For instance, a discrete input space could easily use the Levenshtein or Manhattan distance.
With knowledge of the observable space, the distance in a latent space from a variational autoencoder has been used in early implementations of the SVF.
Even without a model of the observable space, more involved functions could be considered such as those from time resolved pair-correlation functions~\cite{konstantinova2022_PRR} or topological data analysis~\cite{koplik2022topological}. 

In the following we made use of Bayesian optimization over a Gaussian process (GP) surrogate model of SVF. 
The GP used a Matern kernel with homoskedastic noise to construct a probabilistic model of $U$ in all input space,  including regions where there are no observations. 
When conditioning the GP, we scaled the value of $U$ to $U\in [0, 1]$. 
We used the Expected Improvement (EI)~\cite{mockus1978application} and Upper Confidence Bound (UCB)~\cite{srinivas2009gaussian} acquisition functions for Bayesian optimization.
The UCB functions presented use a weighting for variance of $\beta$=10, although similar results were obtained for values of $\beta$ ranging from 10 to 100. 
\par

We benchmarked this against a common experimental design of measuring over an optimal grid given allotted resources (e.g., time or number of measurements), described here as grid search.
We note that despite the specific choices used in this work, the extensibility and flexibility of the SVF allows it to be used with any optimization protocol or probabilistic model that can approximate it. 
Herein, we call the SVF modeling procedure used in tandem with the tools of Bayesian optimization (or optimization in general) the Scientific Value Agent (SVA).

\subsection{Characterizing a one-dimensional phase space with simulated X-ray diffraction} \label{subsec:xrd1dim}

\begin{figure}[htbp!]
    \centering
    \includegraphics[width=\columnwidth]{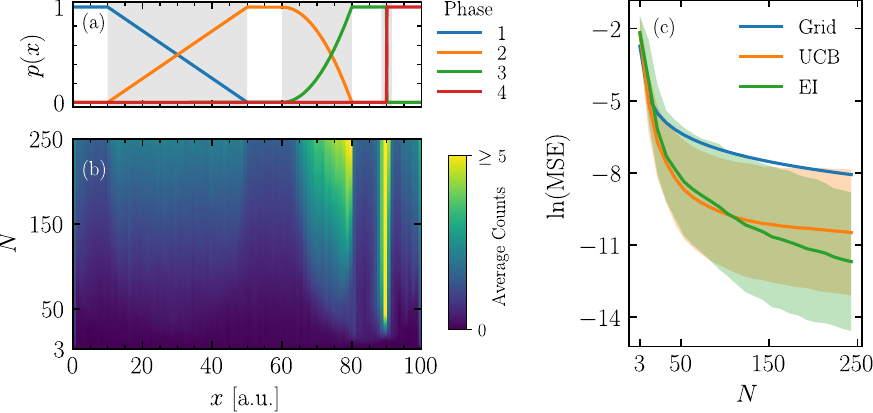}
    \caption{
    Visual summary of the simulated XRD experiment. (a) The proportion of each of the 4 phases as a function of position, $x.$ Regions of change are highlighted with a grey background. (b) A histogram of the average number of counts/experiment as a function of position $x$ and the current dataset size $N$ using the UCB acquisition function with $\beta=10.$ Results are averaged over 300 independent experiments. (c) The average value of the natural log of MSE as a function of $N,$ plotted with a confidence interval of 2 standard deviations. \label{fig:xrd1dim}
    }
\end{figure}

We first tested the SVA \emph{in silico} using the simulated X-ray diffraction measurement of a library that contained linear mixtures of four phases [Fig.~\ref{fig:xrd1dim}]. 
This sampling of a one-dimensional space is common in studying phase behavior over composition or state variables \cite{maffettone2023self, Hua_2021, Olds_2017}.
The four XRD patterns corresponding to the phases were defined by a series of randomly placed Gaussian peaks over a constant background.
Normally distributed noise was introduced to the observation at each query [Fig.~S1]. 
In order to simulate reasonable and challenging types of phase changes, we chose functional forms to represent first and second order transitions: sigmoidal to approximate a discontinuous first order transition, and linear and quadratic for different rates of second order.
These are highlighted in the three shaded regions of Figure~\ref{fig:xrd1dim}(a). 
In an optimal measurement of this compositional library, sampling density should be correlated to the rates of change of the phases.
\par

To quantify sampling performance, we tracked the mean squared error (MSE) between the true observation space (phase fractions) and the observation space that could be reconstructed by the sampled queries (assuming oracle knowledge of phase fractions given the observable). 
The reconstructed dataset is produced by linearly interpolating observations between measured points. 
As shown in Figure~\ref{fig:xrd1dim}(c), this metric will decay as more observations are made, with a smaller error corresponding to more robust sampling. 

Figure~\ref{fig:xrd1dim}(b) shows the average sampling histogram of the SVA over 300 independent experiments. 
Even at small $N,$ we see that the three phase boundary regions were sampled in proportions commensurate with the rate of change of the phases in those regions. 
The linear change region was only sparsely sampled (but still sampled more compared to regions of no change), whereas the quadratic region was sampled much more densely. 
Nonetheless, the region of near-instantaneous change was sampled most densely and earliest, as the algorithm discovered  this  very sharp boundary, and therefore required more samples to produce an accurate representation of the observable in that region.

Both choices of acquisition function outperformed the optimal grid design  by roughly an order of magnitude. 
This performance was apparent in the low-$N$ and limiting cases. 
Not only did the SVA procedures propose experiments in relevant regions of space, they also modeled regions of significant change more efficiently than conventional methods.
\par

We also considered using Bayesian inference as a DOE benchmark. This approach first clustered the data, then trained a probabilistic regressor to predict the cluster labels, and finally queried new points where the uncertainty was maximized \cite{Kusne_2023_Matter}.  
We found this methodology to be strongly dependent on the chosen number of clusters, not necessarily more performant than a grid search in the limiting case, and less performant on low N [Fig.~S5]. 
While Bayesian inference can be a powerful tool \cite{Kusne_2023_Matter}, it is intrinsically model dependent, increasing in potency when a more accurate model for the system is available. 
In Bayesian inference, a label assignment is necessary (here accomplished by K-means clustering).  SVA doesn't require a model or labels, but can incorporate one through the correlation function, $g(\vec y_i, \vec y_j)$.
Considering this comparison and the prevalence of grid searches at actual beamlines, we chose the grid search technique as a our benchmark.

\subsection{Characterizing a two-dimensional space with a periodic interface} \label{subsec:sine2phase}
\begin{figure}[htbp!]
    \centering
    \includegraphics[width=\columnwidth]{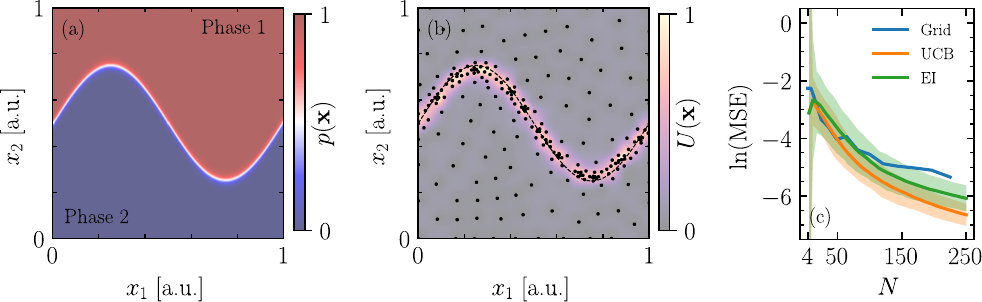}
    \caption{
    Visual summary of the 2-phase, 2-dimensional sinusoidal interface experiment. (a) The two phases, shown in red and blue. The interface, where the phases are in equal contribution, is white. (b) The SVF, approximated by a GP and scaled to values between 0 and 1, shown as a backdrop to the sampling results of $N=250$ points using the UCB acquisition function with $\beta=10.$ The phase boundary is shown as a dotted line. (c) The average value of the natural log of MSE as a function of $N,$ plotted with a confidence interval of 2 standard deviations (smaller is better). A total of 300 experiments over random initial points and model/optimizer seeds were performed for experiment statistics. \label{fig:sine2phase}
    }
\end{figure}

We completed a second \emph{in silico} test, which sought to characterize a two-dimensional library of sample compositions defined by coordinates $\vecx \in \mathbb{R}^2.$ 
In this case, the library contained only two phases, separated by a sharp periodic boundary [Fig.~\ref{fig:sine2phase}(a)]. 
The observation of phases and their mixtures was characterized by a spectrum, simulated using noisy Gaussian functions centered at two different locations in space [Fig.~S2]. 
As above, phases were linearly mixed, with the proportion of the phases given by a sigmoid function of the position on the wafer,
\begin{equation} \label{eq-prop-sinusoid}
    \begin{split}
    \quad b(x_1) = \frac{1}{2} + \frac{1}{4} \sin(2\pi x_1),
    \\
    p(\vecx) = \frac{1}{1 + \exp{-50 [x_2 - b(x_1)]}}.
    \end{split}
\end{equation}
Designed to be a drastic and challenging test, the resultant phase-dependence on position can be see in Figure~\ref{fig:sine2phase}(a).
\par

We compared the performance of the SVA against conventional methods using the same metrics from above. 
Again it was clear that an active learning approach coupled with SVF outperforms conventional measurement techniques [Fig.~\ref{fig:sine2phase}(c)]. 
Additionally, we examined how the SVA queried the space around the phase boundary.
Even in data-limited conditions, the approach successfully mapped out regions of significant change, while still sufficiently sampling relatively constant regions of phase space.
As shown in Figure~\ref{fig:sine2phase}(b), the sampling focused on the most information rich region,  highlighted around the curve  $b(x_1).$ 

Compared to the conventional grid search benchmark, the SVA outperformed this baseline by roughly an order of magnitude.
The results of a single SVA experiment using the UCB acquisition function with a total of 250 samples show a dense sampling of the interface, without under-sampling the surrounding area.
Both UCB and EI behaved comparably and outperformed the baseline, with UCB being more exploitative of the narrow transition region.
Although trade-offs with acquisition functions are expected, optimizing over the SVF was robust regardless of acquisition function choice.

\subsection{Characterizing the subtle phase transitions of Barium Titanate}
The final example we present in the active setting highlights the case where naive data-driven approaches fail~\cite{maffettone2021constrained}.
Furthermore, we used this to demonstrate the capacity to integrate more physics aware correlation functions into the SVF to improve the expressiveness of the surrogate modeling.
We emulated a continuous valued experiment in which total scattering data of BaTiO$_3$ were measured as a function of temperature, by interpolating a dataset measured over 5\textdegree\,C intervals at the Pair Distribution Function beamline at the National Synchrotron Lightsource II [Fig.~S3].
These data contains incredibly subtle transitions between four distinct crystallographic phases (rhombohedral, orthorhombic, tetrahedral, and cubic)
that are difficult to distinguish using data-driven approaches~\cite{maffettone2021constrained}.
Using established methods, we trained an ensemble of convolutional neural networks to predict these phases from simulated diffraction patterns, and used the trained models to create an encoding of the noisy experimental data~\cite{Maffettone_XCA}.
We used the SVA procedure to created a surrogate model for a SVF that used these encodings to calculate the observation space correlation function, $g(\vec y_i, \vec y_j)$.
\par

\begin{figure}[htbp!]
    \centering
    \includegraphics[width=\columnwidth]{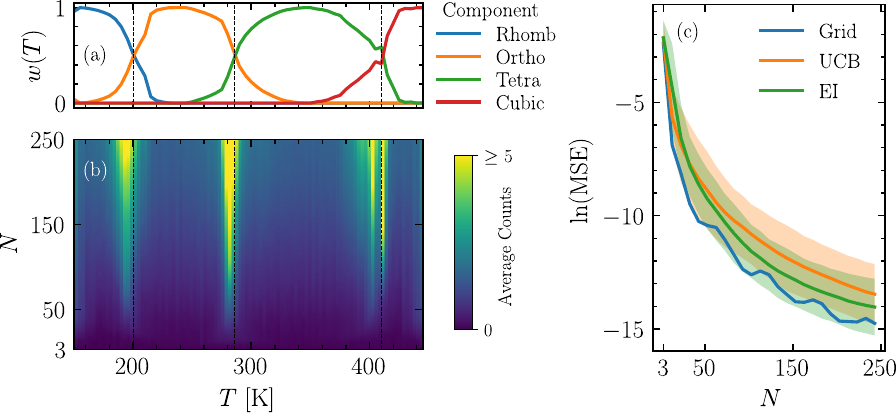}
    \caption{
      Results of the  BaTiO$_3$ experiment.
      (a) Phase fractions determined from Rietveld refinement. Refinement results applying Gibbs phase rule to normalized R$_{wp}$ and renormalizing onto $[0,1]$ (SI).  The dotted lines show the detected phase transitions according to the cryostream gas temperature, and will differ by a lag from the sample temperature. (b) A 2-dimensional histogram of the average number of queries as a function of temperature $T$ and current dataset size using the EI acquisition function. Results are averaged over 300 independent experiments. (c) The average value of the natural log of MSE as a function of $N,$ plotted with a confidence interval of 2 standard deviations. \label{fig:bto}
    }
\end{figure}

Following the same procedure as the previous examples, we showcase the results of the sampling as a function of the number of queries in Figure~\ref{fig:bto}(b).
The SVA correctly identified and attended to the phase transitions extracted by data refinement.
We compared approaches by considering the ability of the resultant dataset to construct the Rietveld refined compositions [Fig.~\ref{fig:bto}(a)].
Without the inclusion of a deep learned embedding, the SVA produced datasets on par with conventional methods; however, by combining the flexibility of the SVF with a deep learned embedding, it autonomously up-sampled the phase changes of BaTiO$_3$ [Fig.~S4].

Because total scattering is a measure of bulk state, it captures more phase coexistence than is present locally throughout the sample, and the first-order phase transitions in BaTiO$_3$ appear gradual and continuous.
Subsequently, a grid search approach could reconstruct the bulk compositions from Rietveld refinements as well as, if not better than, the SVA approach  [Fig.~\ref{fig:bto}(c)].
Nonetheless, grid search methods failed to focus on the unique physical behavior of the transitions, which would be exposed through pair distribution function or spectroscopic analysis.   
This highlights the potential for the SVA to suggest clarifying experiments in the multifidelity or multimodal setting~\cite{maffettone2023self}.

\subsection{In-line analysis of nanoparticle synthesis}

In the previous three examples, we demonstrated the use of the SVF in an adaptive setting across diverse problems relevant to materials science. 
We acknowledge that a variety of algorithms---or none at all---may be preferable for driving an experiment.
Therefore, we highlighted the breadth of the approach by applying the SVF in a passive analysis setting.
We deployed the SVF to visualize a spectroscopy dataset produced  during an automated nanoparticle synthesis experiment at the Center for Functional Nanomaterials at Brookhaven National Laboratory.
The  dataset consisted of $N=375$ experimental flow reactor conditions ($\vecx_i$) and the corresponding UV-vis absorption spectrum ($\vecy_i$). 
The nanoparticle synthesis experiments were performed in a flow reactor by varying 4 experimental parameters: the volume of sodium citrate (NaCit, 16~mmol/L), Chloroauric acid (HAuCl$_4,$ 2~mmol/L), hydrochloric acid (HCl, 10~mmol/L), and sodium hydroxide (NaOH, 10~mmol/L). 
The total volume of liquid in any experiment is always equal to 40 $\mu$L, (the size of the droplet in the flow reactor). 
This reduces the number of degrees of freedom to 3, wherein HCl and NaOH are used to drive the reaction pH.
UV-vis absorption spectra were then taken of the final reaction products. 
\par

\begin{figure}[htbp!]
    \centering
    \includegraphics[width=0.9\columnwidth]{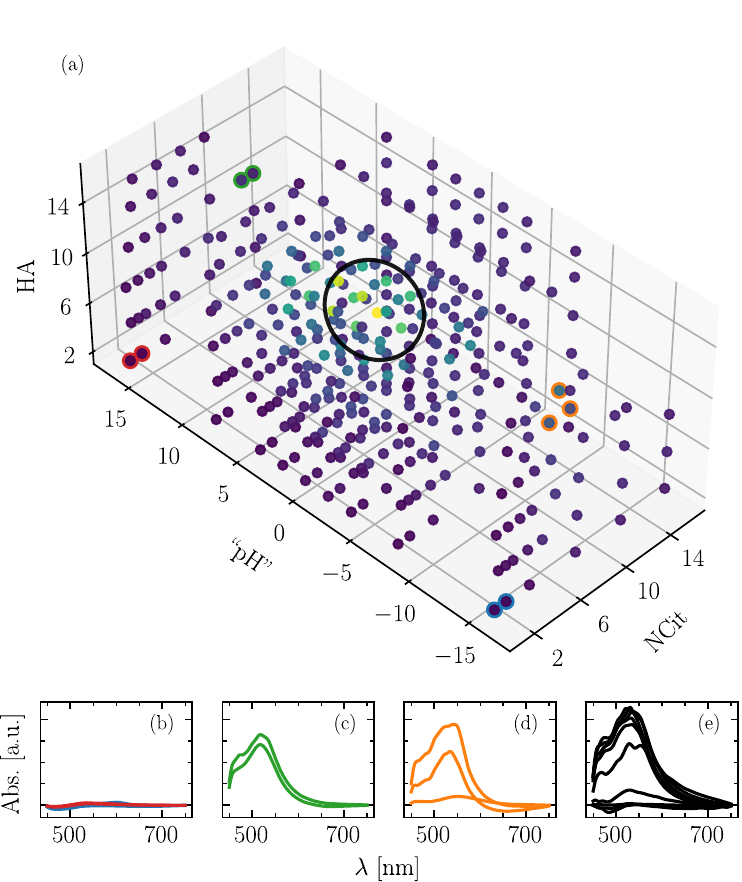}
    \caption{
    Visual summary of the UV-vis experiment. (Top; a) The input reactor conditions. Colors are given by the scientific value (scaled to $[0, 1],$ purple to yellow). (Bottom) Spectra corresponding to the colored outlined points in the top plot, shown in order of increasing value as defined by the average SVF in that cluster of points. (b) The red and blue points show the lowest value ($0.01$) around static regions of experiment space that produce non-absorbing particles. Increasing patches of scientific value (ranges shown) led the scientist to find (c) green regions ($0.09$ to $0.13$) with stable and significant absorbance and (d) regions of gradual change in absorbance in orange ($0.23$ to $0.35$). Lastly, the SVF analysis shown in black highlights a region of dramatic local change ($0.49$ to $0.89$). This cluster was selected by taking the point of highest value and finding its 10 nearest neighbors.\label{fig:uv}
    }
\end{figure}

The experiments were performed via domain experts using a grid search with manual intervention. 
To assist in processing the large dataset of measurements, $\mathcal{D}_N,$, we computed $U(\vecx_i, \vecy_i, \mathcal{D}_N)$ for all $(\vecx_i, \vecy_i) \in \mathcal{D}_N,$.
This offered the scientists a visual representation of scientific value [Fig.~\ref{fig:uv}(a)].
\par

The analysis provided by the SVF equipped the scientist with an actionable visualization tool that keeps the human in the loop of an otherwise automated experiment.
The visualization highlighted regions of high scientific value, as the dataset was growing, by linking the positions in phase space to their individual spectra.
By cross referencing the regions of high value with their respective spectra, the scientist was able to examine the SVF analysis and validate it with their own insight. 
We expect the combination of this advancement with user interface engineering will create a potent tool that impacts a variety of analysis techniques. 

\section{Conclusion}

In this work, we presented the Scientific Value Function, which replicates the judgement of human experts, and recasts a dataset of higher dimensional observables into scalar measures of value. 
By quantifying value, the SVF creates an epistemic research objective that can be optimized without the need for feature engineering or a scalar observable.
We demonstrated the deployment of the SVF in an adaptive learning context (SVA), how it can be complemented by ML or data reduction techniques, and how it can be used in a streaming analysis deployment for visualizations that accelerate decision making by the human expert. 
Because the approach provides a flexible, experiment-agnostic path for building autonomous workflows, it has far reaching implications for accelerated science across physics, chemistry, materials, and biology.

\begin{acknowledgements}
This research is supported in part by Brookhaven National Laboratory (BNL), Laboratory Directed Research and Development (LDRD) Grants: No. 22-059, ``Precision synthesis of multiscale nanomaterials through AI-guided robotics for advanced catalysts,” and No 23-039, ``Extensible robotic beamline scientist for self-driving total scattering studies." This research also used resources of the Center for Functional Nanomaterials, the PDF (28-ID-1) Beamline and resources of the National Synchrotron Light Source II, which are U.S. Department of Energy Office of Science User Facilities, at Brookhaven National Laboratory under Contract No. DE-SC0012704.
\end{acknowledgements}


\section*{Declaration of interests}
The authors declare no competing interests, financial or otherwise.

\section*{Data statement}

All software and data used to generate the results in this manuscript can be found open source under a BSD-3-clause license~\href{https://github.com/matthewcarbone/ScientificValueAgent}{github.com/matthewcarbone/ScientificValueAgent}. Tarballs containing all of the data used to generate our results are hosted open access at \href{https://doi.org/10.5281/zenodo.8184819}{doi.org/10.5281/zenodo.8184819}. All figures presented in this manuscript can be regenerated by using these data, and the notebooks stored at the GitHub link above.


\bibliography{bibliography.bib}

\end{document}


\title{Emulating Expert Insight: \\ A Robust Strategy for Optimal Experimental Design\\ Supplementary Information}

\author{Matthew R.~Carbone}
\email{mcarbone@bnl.gov}
\affiliation{Computational Science Initiative, Brookhaven National Laboratory, Upton, NY 11973, USA}

\author{Hyeong Jin Kim}
\affiliation{Center for Functional Nanomaterials, Brookhaven National Laboratory, Upton, NY 11973, USA}

\author{Shinjae Yoo}
\affiliation{Computational Science Initiative, Brookhaven National Laboratory, Upton, NY 11973, USA}

\author{Daniel Olds}
\affiliation{National Synchroton Light Source II, Brookhaven National Laboratory, Upton, NY 11973, USA}

\author{Howie Joress}
\affiliation{Materials Measurement Science Division, National Institute of Standards and Technology, Gaithersburg, MD 20899, USA}

\author{Brian DeCost}
\affiliation{Materials Measurement Science Division, National Institute of Standards and Technology, Gaithersburg, MD 20899, USA}

\author{Bruce Ravel}
\affiliation{Materials Measurement Science Division, National Institute of Standards and Technology, Gaithersburg, MD 20899, USA}

\author{Yugang Zhang}
\email{yuzhang@bnl.gov}
\affiliation{Center for Functional Nanomaterials, Brookhaven National Laboratory, Upton, NY 11973, USA}

\author{Phillip M.~Maffettone}
\email{pmaffetto@bnl.gov}
\affiliation{National Synchroton Light Source II, Brookhaven National Laboratory, Upton, NY 11973, USA}

\date{\today}

\maketitle
\clearpage
\pagebreak
\widetext
\begin{center}
\textbf{\large Supplemental Material: The Scientific Value Agent}
\end{center}

\setcounter{equation}{0}
\setcounter{section}{0}
\setcounter{figure}{0}
\setcounter{table}{0}
\setcounter{page}{1}
\makeatletter
\renewcommand{\thesection}{S\arabic{section}}
\renewcommand{\thesubsection}{S\arabic{section}.\arabic{subsection}}
\renewcommand{\theequation}{S\arabic{equation}}
\renewcommand{\thefigure}{S\arabic{figure}}

\begin{figure}[htb!]
    \centering
    \includegraphics{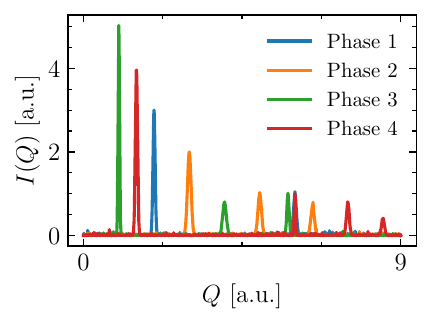}
    \caption{Example synthetic diffraction patterns of the four phases explored in characterizing a one-dimensional phase space with simulated X-ray diffraction.}
    \label{fig:s-simxrd}
\end{figure}

\begin{figure}[htb!]
    \centering
   \includegraphics{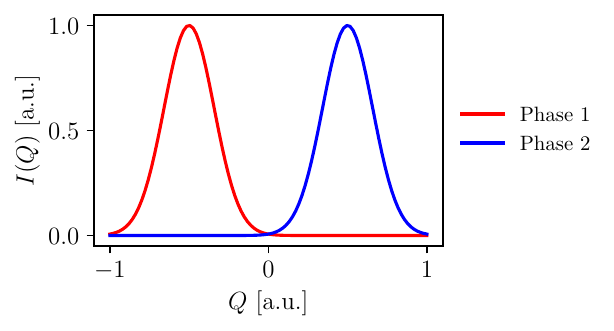}
    \caption{Example synthetic spectral data of the two phases explored in characterizing a two-dimensional space with a periodic interface.}
    \label{fig:s-sim2d}
\end{figure}

\begin{figure}[htb!]
    \centering
    \includegraphics{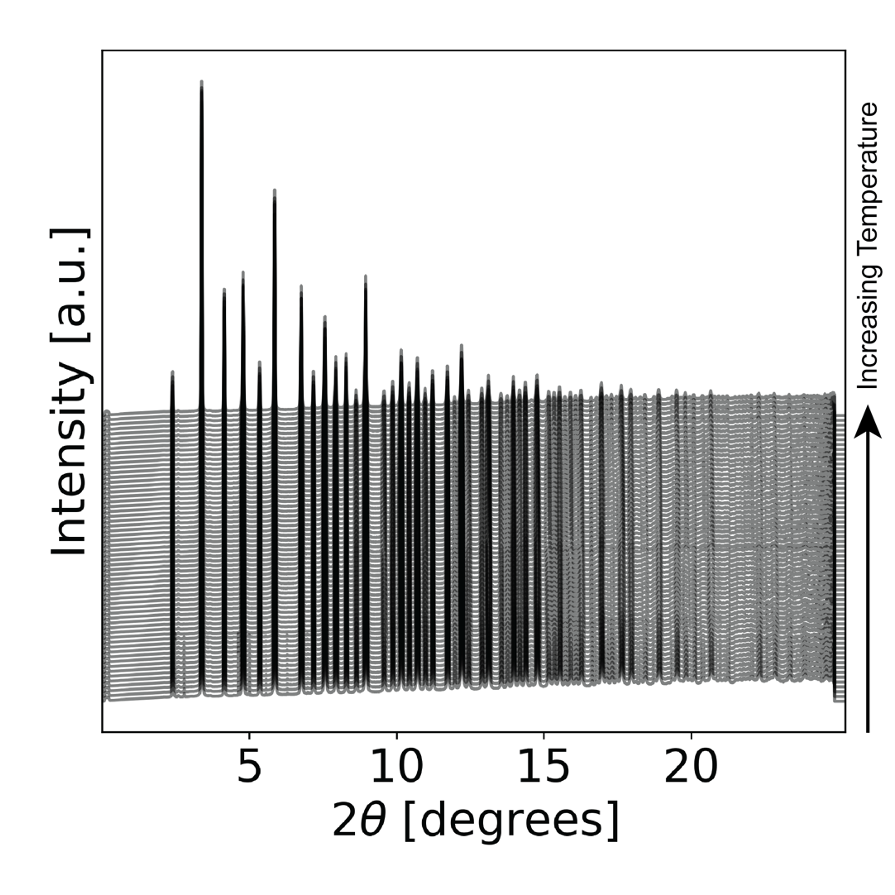}
    \caption{Experimentally measured BaTiO$_3$ dataset over the temperature range 150 to 450K, shows three phase transitions that are imperceptible even to conventional refinement techniques.}
    \label{fig:s-bto-raw}
\end{figure}

\begin{figure}[htb!]
    \centering
   \includegraphics{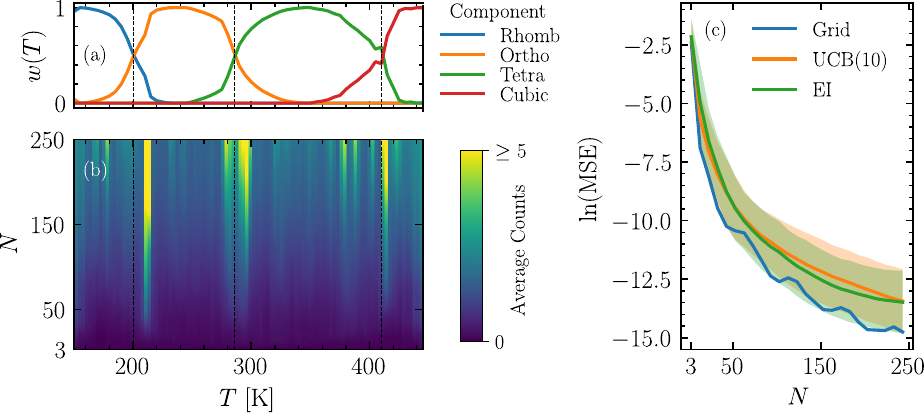}
    \caption{Results of the  BaTiO$_3$ experiment that deploys SVA using only Euclidian distance as the observable correlation function.
      (a) Phase fractions determined from Rietveld refinement. Refinement results applying Gibbs phase rule to normalized R$_{wp}$ and renormalizing onto $[0,1].$ The dotted lines show the detected phase transitions according to the cryostream gas temperature, and will differ by a lag from the sample temperature. (b) A 2-dimensional histogram of the average number of queries as a function of temperature $T$ and current dataset size using the UCB acquisition function. Results are averaged over 300 independent experiments. (c) The average value of the natural log of MSE as a function of $N,$ plotted with a confidence interval of 2 standard deviations.  }
    \label{fig:s-bto-no-deep}
\end{figure}

\begin{figure}[htb!]
    \centering
    \includegraphics{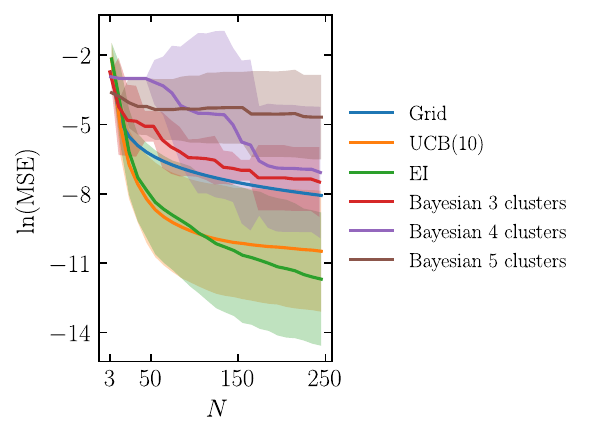}
    \caption{Results comparing the MSE from Bayesian inference over clustering techniques to SVA. For the simulated XRD experiment with 4 underlying components we clustered these data as they were obtained using K-means, $k={3,4,5}$, trained a logistic regressor to predict the cluster labels, and sampled according to the uncertainty of the trained model to choose the next experiment. Each campaign was repeated 10 times to collect statistics, with 2 standard deviations shown as confidence intervals.}
    \label{fig:s-bayesian-clustering}
\end{figure}

\section{Calculation of crystallographic residual}

Normalized weighted residue ($\text{norm.\,} R_{wp}$) were calculated for each fit of a refined phase by extracting the calculated $R_{wp}$ from the Rietveld refinements using the TOPAS software \cite{coelho2018topas}, and then normalizing by the average $R_{wp}$ for each refined phase, and such that the result spans 0 to 1 for visual clarity.

\begin{equation}
    \text{norm.\,} R_{wp} = \frac{(R_{wp} - <R_{wp}>) - \min(R_{wp} - <R_{wp}>)}{\max(R_{wp} - <R_{wp}> - \min(R_{wp}-<R_{wp}>))}
\end{equation}

In this description, a normalized  $R_{wp}$  of 0 would correspond to a phase existence. Subsequent re-normalization was accomplished by applying Gibb's phase rule, and only allowing for 2 phases with non-zero composition, $1-R_{wp}$, then again normalized on $[0, 1]$ for output compositions. 

\bibliography{bibliography.bib}